\documentclass[compsoc,twocolumn,10pt,letterpaper]{IEEEtran}
\usepackage[cmex10]{amsmath}
\usepackage{graphicx,array,tabularx}
\begin{document}
%
\title{Metabolic Flux Analysis in Isotope Labeling Experiments using the Adjoint Approach}
%
%
%
\author{St\'ephane~Mottelet,
         Gil~Gaullier
        and~Georges~Sadaka%
\thanks{S. Mottelet  is with the Department
EA4297 TIMR, Sorbonne University, Universit\'e de technologie de Compi\`egne, rue du docteur Schweitzer, F-60203 Compi\`egne, France e-mail: stephane.mottelet@utc.fr.}%
\thanks{G. Gaullier is with the  Department EA2222 LMAC, Sorbonne University, Universit\'e de technologie de Compi\`egne.}%
\thanks{G. Sadaka is with the Department CNRS, UMR7352 LAMFA, Universit\'e de Picardie Jules Verne.}%
\thanks{Manuscript received xxxx 00, 2015; revised xxxx 00, 201x.}}

%
\maketitle
\begin{abstract}
Comprehension of  metabolic pathways is considerably enhanced by metabolic flux analysis (MFA-ILE) in isotope labeling experiments. The balance equations are given by hundreds of algebraic (stationary MFA) or ordinary differential equations (nonstationary MFA), and reducing the number of operations is therefore a crucial part of reducing the computation cost. The main bottleneck for deterministic algorithms is the computation of derivatives, particularly for nonstationary MFA. In this article we explain how the overall identification process may be speeded up by using the adjoint approach to compute the gradient of the residual sum of squares. The proposed approach shows significant improvements in terms of complexity and computation time when it is compared with the usual (direct) approach. Numerical results are obtained for the central metabolic pathways of \emph{Escherichia coli} and are validated against reference software in the stationary case. The methods and algorithms described in this paper are included in the \texttt{sysmetab} software package distributed under an Open Source license at http://forge.scilab.org/index.php/p/sysmetab/.
\end{abstract}


%
\IEEEpeerreviewmaketitle

\section{Introduction}
\IEEEPARstart{N}{onstationary} MFA can contribute significantly to the comprehension, both qualitative and quantitative, of a metabolic network~\cite{wiechert2013NMFA}, but before the creation of the Elementary Metabolites Unit (EMU) framework \cite{young2008elementary,young2014inca} the high computation cost of nonstationary MFA prevented the development of efficient software. However, topological network reduction methods (such as the one that allows the EMU framework) are not the only way of curbing the computation cost of nonstationary MFA. Given that both stationary and nonstationary MFA belong to the class of general inverse problems where measurements depend on parameters through a state equation, there is a technique, namely the adjoint approach \cite{lionsmagenes,chavent,plessix}, that may be used to greatly improve the computation of different kinds of derivatives.

The main contribution of this paper is demonstrating that independently of the topological network reduction, the adjoint approach considerably speeds up the computation of the gradient residual sum of squares (RSS), meaning that the entire process of estimating unknown parameters is accelerated significantly. We show that in the (usual) direct approach the sum of the times required for the state and gradient calculations is dominated by the computation time of the gradient itself, whereas in the adjoint approach the two computation times are of the same order. While the direct approach has a cost that is unavoidably proportional to the number~$p$ of unknown parameters (fluxes and pool sizes), in the case of the adjoint approach the cost of the gradient computation is independent of~$p$. The algorithms described below are implemented in the accompanying Open Source software \texttt{sysmetab}, which can handle stationary and nonstationary MFA.

As theoretical results are asymptotical and can be degraded by practical implementation bottlenecks, we perform a time analysis for both gradient  computations on the central metabolism of \emph{E. Coli}, in which we compare overall computation times for the direct and adjoint approaches with respect to labeling state sizes: reduced vs. full cumomer set and stationary vs. nonstationary data. We contrast estimated flux values for the stationary data with the values obtained from
the reference softwares \verb&13CFLux2& \cite{13CFLUX2} and \verb1influx_s1 \cite{influxs}, which do not implement the adjoint approach. We detail the distribution of running time on the network mentioned above and on a larger, poorly defined version of this network.  We show that in this particular context, when multiple flux estimations from different perturbed data are made repeatedly (i.e. the Monte Carlo method is applied), \verb1sysmetab1 is competitive with other software. For parameter estimations using time-dependent labeling data generated by stationary flux estimations and realistic pool size values, the ratio of running times from one approach to the other exceeds 20.

The paper is organized as follows: In Section~\ref{sec:math}, we introduce the mathematical structure of $^{13}$C MFA.
Mathematical aspects of the algorithms implemented in  \texttt{sysmetab} are given in Sections \ref{sec:MFA} and \ref{sec:NMFA} respectively for stationary and nonstationary MFA, and their complexity is compared. Implementation details are described in Section \ref{sec:impl}. Section \ref{sec:num} presents numerical results obtained by \verb1sysmetab1.
\section{Optimization model of MFA}
\label{sec:math}
The aim of $^{13}\mathrm{C}$ metabolic flux analysis is to determine fluxes $v$, pool sizes ${m} $ (in the nonstationary case), expressed as a vector of $p$ parameters $\theta=(v,{m})$, such that experimental labeling data $y$ and other measurements $w$, e.g. input/output fluxes and pool sizes, are best fitted to their simulated values. This means minimizing the objective function
\begin{equation}
J(\theta) =\textstyle\frac{1}{2}\Vert g(x(\theta) ,\theta)  -y\Vert_\mathcal{Y} ^2+\textstyle\frac{1}{2}\Vert h(\theta) -w\Vert_\mathcal{W} ^2,
\label{eq:data}
\end{equation}
where $g$ and $h$ are given functions and $x(\theta) $ denotes the (possibly time-dependent) $n$ labeling states of the metabolic network, implicitly defined  by the state equation
\begin{equation}
f(x,\theta) =0.
\label{eq:state}
\end{equation}
This equation takes the form of a system of algebraic or ordinary differential equations, respectively for stationary or nonstationary metabolic flux analysis. The notations $\|\cdot\|_\mathcal{Y} $ and $\|\cdot\|_\mathcal{W} $ are used for the norms in the space of labeling and non-labeling measurements, denoted by $\mathcal{Y}$ and $\mathcal{W}$ respectively. These norms (may) model variations due to experimental noise by introducing weighting covariance matrices. In both the stationary and nonstationary cases the spaces $\mathcal{Y}$ and $\mathcal{W}$ are finite-dimensional normed vector spaces, hence $y$ and $w$ can be considered as vectors containing real measured data.

Below, $\Theta$ denotes the space of admissible parameters, i.e. the space describing realistic physical parameters comprising constraints on fluxes, such as  the stoichiometric equations for fluxes. In order to find
\begin{equation}
\hat \theta=\arg\min_{\theta\in{\Theta}} J(\theta) ,
\label{eq:opt}
\end{equation}
different optimization methods can be considered. Finding $\hat \theta$ can be performed by determining the solution of successive linearized problems, as the in Gauss-Newton method (GN) with a first-order Taylor expansion of $g(x(\theta) )$ with respect to $\theta$ by the computation of the sensitivity matrix $x'(\theta) $. In the nonstationary case this $n\times p$ matrix is time-dependent and satisfies a system of differential equations, which means that it is very expensive to compute.

In contrast, if the minimization is considered as a general nonlinear optimization problem, only the gradient of $J$ needs to be computed, as in the Broyden-Fletcher-Goldfarb-Shanno (BFGS) and Sequential Quadratic Problem (SQP) methods. In the 1970s the adjoint state method was developed to compute the gradient in the situation frequently encountered in control theory where the function to minimize depends indirectly on parameters through state variables, with a minimal cost~\cite{lionsmagenes,chavent,plessix}. This cost is asymptotically independent of the number of parameters $p$, and is equivalent to the cost of a single-state equation, while using the sensitivity matrix generates a cost proportional to $p$.
\section{Stationary MFA}
\label{sec:MFA}
The cumomer fraction variables~$x=(x_1,\ldots,x_{N})$ model introduced in~\cite{wiechert1} means that the structure of balance equations in stationary MFA is given by a cascade  of linear algebraic equations. If $x_k$ denotes the vector of cumomer fractions of weight $k$ ,  $x_k^{\leftarrow}$ the vector of input cumomers of weight $k$, and $x_{<k}$, $x_{\leq k}$ the sequences $(x_i)_{i<k}$ and $(x_i)_{i\leq k}$, respectively, then the state equation~\eqref{eq:state} can be written as
\begin{equation}
f_k(v,x_{\leq k},x_k^{\leftarrow})=0,\quad 1\leq k\leq N,
\label{eq:cascade}
\end{equation}
where the function $f_k$ is affine with respect to $x_k$ as
\begin{equation*}
f_k(v,x_{\leq k},x_{\leq k}^{\leftarrow})=A_k(v)x_k+b_k(v,x_{<k},x_{\leq k}^{\leftarrow}),\,
\end{equation*}
the  matrix $A_k$ and vector $b_k$ being determined by the structure of the metabolic network under consideration. From the forward cascade structure of \eqref{eq:cascade}, we successively obtain $x_1$, $x_2$,\dots,$x_N$ by solving at each step a system of linear equations. For the sake of conciseness we do not consider the terms which explicitly depend on $\theta$ in \eqref{eq:data}, but only labeling measurements. In this case, measurements linearly depend on the state variable, so that the functional to minimize is of the form
$$
J(v)=\textstyle\frac{1}{2}\Vert Cx(v)-y\Vert^2,
$$
for a given matrix $C=(C_1,\dots,C_N)$ depending on the measurement model.
\subsection{Direct approach}
For a given flux vector $v$, we first need to solve the state equation, which provides the cumomer vector $x(v)$. The sequence of derivatives $(x'_k)_{1\leq k\leq N}$ is then computed by solving each step of the forward cascade (the sum in the second row vanishes for $k=1$)
\begin{equation}
\left.
\begin{split}\label{eq:jac}
A_k(v)x_k'=&-\partial_vf_k(v,x_{\leq k},x_{\leq k}^{\leftarrow})\\
+&\textstyle\sum_{i<k}\partial_{x_i}b_k(v,x_{< k},x_{\leq k}^{\leftarrow})x_i',\;k\geq 1,\end{split}
\right.
\end{equation}
which is obtained by implicit differentiation of \eqref{eq:cascade} with respect to~$v$. Finally, the gradient of $J$ is computed as
\begin{equation}
\nabla J(v)=\sum_{k=1}^N x'_k(v)^\top C_k^\top(Cx(v)-y) ,
\label{eq:fwdgrad}
\end{equation}
where the symbol $^\top$ denotes vector and matrix transposition.
\subsection{Adjoint approach}
The adjoint approach is based on the Lagrangian functional
\begin{align*}
\mathcal{L}(x,v,\lambda)&= \textstyle\frac{1}{2}\Vert Cx-y\Vert^2 + \sum_{k=1}^N \lambda_k^\top f_k(v,x_{\leq k},x_{\leq k}^{\leftarrow}),
\end{align*}
where $\lambda=(\lambda_k)_{1\leq k\leq N}$ is the sequence of adjoint variables. Expanding the adjoint equation $\partial_x\mathcal{L}(x,v,\lambda)=0$ \cite{plessix} leads to the backward cascade (the sum vanishes for $k=N$)
\begin{equation}
\left.
\begin{split}
\label{eq:adjcascade}
A^\top_k(v) \lambda_k=&\, C_k^\top (Cx-y)\\
-&\textstyle\sum_{i>k}  \partial_{x_k}b_i(v,x_{<i},x_{\leq i}^{\leftarrow})^\top\lambda_i,\;k \leq N.\\[2mm]\end{split}
\right.
\end{equation}
For a given $v$, equations \eqref{eq:adjcascade} are solved backwards, once the  cumomer vector $x(v)$ has been computed from \eqref{eq:cascade}. This procedure provides  successively $\lambda_N,\lambda_{N-1},\dots,\lambda_1$. The gradient $\nabla J$ can then be obtained by the solutions of the $N$ previous linear systems as follows
\begin{equation}
\nabla J(v)=\sum_{k=1}^N  {\partial_v f_k}(v,x_{\leq k},x_{\leq k}^{\leftarrow})^\top \lambda_k.
\label{eq:adjgrad}
\end{equation}
The adjoint approach can also be used to compute the derivative of any function of $x$ with respect to $v$ without requiring $x'(v)$ to be computed. One example of this would be the output sensitivity matrix  $S(v)=Cx'(v)$, which is used in first order sensitivity analysis \cite{cintron2009sensitivity}. This matrix is also used at each iteration of the optimization method to compute the  descent direction when a Gauss-Newton type algorithm is used \cite{influxs}). In both usages, the adjoint approach allows $S(v)$ to be computed as a fraction $n_y/n$ of the usual cost, where $n_y$ is the number of measurements and $n$ the size of cumomer vector $x$.
\subsection{Computational comparison}
Both approaches involve the computation of the state $x(v)$ in the computation of $\nabla J(v)$. To achieve this, the matrices $A_k(v)$ need to be factorized and the derivatives $ {\partial_v f_k}$ and $\partial_{x_i}b_k$ computed for $i<k$. The crucial difference between the two approaches is to be found in the two remaining equations: for a given~$k$, equation~\eqref{eq:jac} involves solving a linear system whose right-hand side is a matrix with $p$ columns, whereas in equation~\eqref{eq:adjcascade} the right-hand side is a (single) column vector. A subsequent analysis of equations \eqref{eq:fwdgrad} and \eqref{eq:adjgrad} shows that even though the elementary operations are identical, the computation cost of equation~\eqref{eq:adjgrad} differs from~\eqref{eq:fwdgrad} in that ~${\partial_v f_k}$ is sparse, unlike ~$x'_k$ which is full. While the difference remains negligible for small metabolic networks, our numerical results show that the larger the network, the more significant the reduction in computation cost.

We show in the following section that for nonstationary MFA the improvements obtained from the adjoint approach are of the same magnitude. But, since the problem is time-dependent, the overall computation cost increases considerably. In order for the problem to become tractable, the adjoint approach is essential.
\section{Nonstationary MFA}
\label{sec:NMFA}
In nonstationary MFA, the balance equations are modeled by a cascade of ordinary differential equations given for $1\leq k \leq N$~by
\begin{equation}
\left\{
\begin{split}
x_k(0)&=0,\\
X_k({m} ) \frac{d}{dt} x_k(t)&=f_k(v,x(t),x_k^{\leftarrow})  ,\,t\in]0,T].
\end{split}
\label{eq:timecasc}
\right.
\end{equation}
To simplify the presentation we will consider that the pool sizes are known. We consider that labeling measurements $y_j$ are made at times $\tau_1<\dots<\tau_M$, where $0<\tau_1$ and $\tau_M<T$, so that the functional to minimize is given by
$$
J(v)=\frac{1}{2}\sum_{j=1}^{M}\Vert Cx(\tau_j;v)-y_j\Vert^2,
$$
where  $x(\tau_j;v)$ is the solution of \eqref{eq:timecasc} at time $\tau_j$.

 Since the cumomer fractions $x_k$ are time-dependent, the space $X$ of the labeling states is no longer a finite-dimensional vector space as in stationary MFA, but an infinite-dimensional functional space $L^2([0;T])$ given by square-integrable functions. The inner product $(\cdot,\cdot)_X$ in $X$ is the usual inner product in $L^2([0;T])$, i.e. $(\phi,\psi)_X=\int_0^T \phi(t)^\top\psi(t) dt$.
\subsection{Direct approach}
As in the stationary case, the direct approach is based on the calculation of the partial derivative $x_k'$ with respect to the flux $v$.
When the sequence $(x_k')_{1\leq k\leq N}$ is time-dependent, this approach involves a cascade of ordinary differential equations (ODE)
\begin{equation}
\left.
\begin{split}
X_k({m} ) \frac{d}{dt} x_k'=&\partial_v f_k(v,x_{\leq k},x_{\leq k}^{\leftarrow})+A_k(v)x_k'\\
&+\textstyle\sum_{i<k}  \partial_{x_i}f_k(v,x_{\leq i},x_{\leq i}^{\leftarrow}) x_i',
\end{split}
\right.
\label{eq:timedirect}
\end{equation}
for $1\leq k\leq N$, obtained by the differentiation of \eqref{eq:timecasc} with respect to the flux parameter $v$. In \eqref{eq:timedirect}, the sum in the second row vanishes for $k=1$, and $\frac{d}{dt} x_k'$ denotes the time derivative of $x_k'$. The initial condition related to \eqref{eq:timedirect} is given, for all $k$ such that $1\leq k\leq N$, by $x_k'(0)=0$. The gradient of $J$ is expressed as
\begin{equation}
\nabla J(v)=\sum_{j=1}^M\sum_{k=1}^N x'_k(\tau_j;v)^\top C_k^\top(Cx(\tau_j;v)-y_j) .
\label{eq:timefwdgrad}
\end{equation}
\subsection{Adjoint approach}
In the adjoint approach, the $N$ ODEs \eqref{eq:timecasc} are incorporated into a Lagrangian functional as follows
\begin{align*}
\mathcal{L}(x,p,\lambda)&=\textstyle\frac{1}{2}\sum_{j=1}^M\left\Vert Cx(\tau_j)-y_j\right\Vert^2+\\
+\sum_{k=1}^N&\int_0^T \lambda_k^\top(t)\left(f_k(v,x_{\leq k}(t),x_{\leq k}^{\leftarrow})-X_k({m} ) \textstyle\frac{d}{dt} x_k(t)\right)dt.
\end{align*}
The adjoint equation is obtained by expanding $\partial_x\mathcal{L}(x,p,\lambda)=0$, which leads to the following backwards-in-time ODE cascade
\begin{equation}
\left.
\begin{split}
X_k({m}) \frac{d}{dt}\lambda_k&=A_k(v)^\top \lambda_k\\
-&\textstyle\sum_{i>k} \partial_{x_k}b_i(v,x_{<i},x_{\leq i}^{\leftarrow})^\top\lambda_i,
\end{split}
\right.
\label{eq:timeadj}
\end{equation}
for $1\leq k\leq N$, where the second row vanishes for $k=N$ and the equalities read for $t\in[0,T[\setminus \{\tau_j\}_{1\leq j \leq M}$. The final condition is given for each $k$ by $\lambda_k(T)=0$ and the following jump conditions
\begin{equation}
X_k({m})[\lambda_k(\tau_j^+)-\lambda_k(\tau_j^-)]=C_k^\top (Cx(\tau_j)-y_j),
\label{eq:iniadj}
\end{equation}
occur for $j=1\dots M$. From \eqref{eq:timeadj}-\eqref{eq:iniadj} we obtain the functions $\lambda_k(t)$ which enter into the gradient of $J$ (with respect to $v$) as follows
\begin{equation}
\nabla J(v)=\sum_{k=1}^N \int_0^T\partial_v f_k(v,x_{\leq k}(t),x_{\leq k}^{\leftarrow})^\top \lambda_k\,dt.
\label{eq:timeadjgrad}
\end{equation}
\subsection{Discrete time schemes}
Both the direct and the adjoint approaches in nonstationary MFA require state $x(t;v)$ to be computed, and this can be done efficiently using discrete schemes. From a time grid $t_n=nh$,  with $h=T/N_T$ ($N_T+1$ denoting the total number of points sampling the interval $[0;T]$), discrete schemes are based on approximating the integral of the right-hand side of \eqref{eq:timecasc} over $[t_n;t_{n+1}]$. For the sake of simplicity we consider the implicit Euler scheme, but a similar approach can be used with schemes having a greater order, while also taking into account the possible stiffness of the state equation, such as the implicit trapezoidal rule \cite{Mottelet2012RR} or implicit Runge-Kutta schemes \cite{sandu}.

Denoting the approximations $x_k^n\simeq x_k(t_n)$ and $x_{\leq k}^n\simeq x_{\leq k}(t_n)$, the implicit Euler scheme gives the following  discrete version of~\eqref{eq:timecasc}
\begin{equation}
\left\{
\begin{split}
x_k^0&=0,\\
X_k({m}) \left(x_k^{n+1}-x_k^n\right)&=h f_k(v,x_ {\leq k}^{n+1},x_{\leq k}^\leftarrow),
\end{split}
\label{eq:discrcasc}
\right.
\end{equation}
where the second row has to be considered for $0\leq n<N_T$, and the objective function is given by
$$
J(v)=\textstyle\frac{1}{2}\sum_{j=1}^M\Vert Cx^{n(j)} -y_j  \Vert,
$$
where $t_{n(j)}=\tau_j$, i.e. each measurement time is assumed to correspond to a sampling point of the time grid.

In the direct approach, implicit derivation of scheme \eqref{eq:discrcasc} with respect to $v$ and applying the Euler scheme to the continuous time equation \eqref{eq:timedirect} provides the approximated derivatives $x'_k$ at sampling points $t_n$ by solving
\begin{equation}
\left.
\begin{split}
X_k({m} ) (x_k'^{\,n+1}-x_k'^{\,n})=&h A_k(v)x_k'^{\,n+1}+h\partial_v f_k(v,x_{\leq k}^{n+1},x_{\leq k}^{\leftarrow})\\
+h&\textstyle\sum_{i<k}  \partial_{x_i}f_k(v,x_{\leq i}^{n+1},x_{\leq i}^{\leftarrow}) x_i'^{\,n+1},\\[2mm]
\end{split}
\right.
\label{eq:discrdirect}
\end{equation}
combined with the initial condition $x_k'^{\,0}=0$ for $k=1\dots N$. From the time-dependent state sequence and the time-dependent state derivative sequence, the gradient is obtained by
\begin{equation}
\nabla J(v)=\sum_{j=1}^M\sum_{k=1}^N (x_k'^{\,n(j)})^\top C_k^\top(Cx^{n(j)}-y_j) .
\label{eq:discrdirectgrad}
\end{equation}
In the adjoint approach, the chosen discrete scheme generates $N_T$ equalities incorporated into the Lagrangian functional by  $N_T$ discrete Lagrange multiplier $(\lambda_k^n)_{0\leq n<N_T}$. Considering all cumomer fractions of weight $k$, we obtain the following discrete version of the Lagrangian functional corresponding to \eqref{eq:discrcasc}
\begin{equation}
\begin{split}
&\mathcal{L}(x,p,\lambda)=\textstyle\frac{1}{2}\sum_{j=1}^M\left\Vert Cx^{n(j)}-y_j\right\Vert^2+\\
&\sum_{k=1}^N\sum_{n=0}^{N_T-1}(\lambda_k^n)^\top\left(h f_k(v,x_ {\leq k}^{n+1},x_{\leq k}^\leftarrow)-X_k({m} ) \left(x_k^{n+1}-x_k^n\right)\right).
\end{split}
\label{eq:discrlagr}
\end{equation}
It should be emphasized that the discrete version of \eqref{eq:timeadj} must be established by deriving  the discrete Lagrangian \eqref{eq:discrlagr}, and not simply by choosing discrete schemes corresponding to  \eqref{eq:timeadj} and \eqref{eq:timeadjgrad}. The discrete adjoint equation is completely determined once a discrete scheme is chosen for the state equation. For the Euler scheme \eqref{eq:discrcasc} and the Lagrangian \eqref{eq:discrlagr}, the discrete adjoint scheme is thus given for $n=1\dots N_T-1$ by
\begin{equation}
\left.
\begin{split}
&X_k({m})({\lambda_k^n-\lambda_k^{n-1}})= h A_k(v)^\top \lambda_k^{n-1}\\
&-h\sum_{i>k} \partial_{x_k}b_i(v,x_{<i}^n,x_{\leq i}^{\leftarrow})^\top\lambda_i^{n-1}+\delta_n,
\end{split}
\right.
\label{eq:discradjoint}
\end{equation}
with final condition $\lambda_k^{N_T-1}=0$ and
$$
\delta_n=\left\{
\begin{array}{rl}
C_k^\top (Cx^{n}-y_j),&\mbox{if } n=n(j),\\
0,&\mbox{otherwise},
\end{array}
\right.
$$
which determines~$(\lambda_k^n)_{0\leq n<N_T}$ for each weight $k$. Finally, the gradient of $J$ at $v$ is given by
\begin{equation}
\nabla J(v)=h\sum_{k=1}^N \sum_{n=0}^{N_T-1}\partial_v f_k(v,x^{n+1}_{\leq k},x_{\leq k}^{\leftarrow})^\top \lambda_k^n.
\label{eq:discradjgrad}
\end{equation}
\subsection{Computational comparison}
As in the stationary case, the computation of $\nabla J(v)$ in both the direct and the adjoint approaches involves computing the time-dependent state $x$, and the associated cost will depend on the discrete scheme that is chosen. Once again, the difference between the two approaches lies in the cascades \eqref{eq:timedirect} and \eqref{eq:timeadj}: in the direct approach we have $N$ differential equations with $p$-column matrices, whereas in the direct approach the left-hand side of \eqref{eq:timeadj} contains 1-column vectors. Hence, at the very least, we should expect a gain equivalent to the gain obtained in the stationary case.

Another interesting aspect of the adjoint approach appears when the pool sizes are also unknown: in the direct approach equation \eqref{eq:timedirect} has to be implicitly derived with respect to $m$, which leads to a more complex system to express and to solve than \eqref{eq:timedirect}. In contrast, the structure of the system \eqref{eq:timeadj} is independent of the actual unknown parameters in the adjoint approach and only the final step \eqref{eq:timeadjgrad} changes.
\section{Implementation}
\label{sec:impl}
The methods and algorithms described in this paper have been included in the \texttt{sysmetab} software package, distributed under an Open Source license via http://forge.scilab.org/index.php/p/sysmetab/ and available for Linux and MacOSX platforms.

In order to be processed by \verb1sysmetab1, metabolic network description, carbon atom transition map and measurements (stationary or nonstationary Mass Spectrometry (MS) or Nuclear Magnetic Resonance spectroscopy (NMR) data) need to be coded in a plain text XML that respects the FML (Flux Markup Language) format developed for the \texttt{13CFlux2} software package \cite{13CFLUX2} (whereas \texttt{13CFlux2} only handles stationary problems, the FML input format also allows nonstationary data to be described). For stationary data, FTBL files (input format of the texttt{13CFlux} previous version) are also supported, and these are automatically converted to FML format. The \verb1sysmetab1 software parses the input file and generates a flux identification program in the Scilab language \cite{scilabbook} that is specific to the network under consideration. After execution, results are output to a plain text XML file conforming to the Forward Simulation Markup Language developed for the \texttt{13CFlux2} software. This file can easily be converted into other file formats, and the fact that \verb1sysmetab1 is a command-line tool makes scripting and batch processing straightforward.

Using the adjoint approach is not the only innovative feature of the software. Another novel aspect is the technique chosen for generating the code: starting from the original FML file, the Scilab program is entirely generated using XSL transformations (http://www.w3.org/TR/xslt). These  transformations are specified in XSL stylesheets, written in another XML dialect. XSL is very different from other programming languages in use today in that it is a declarative (as opposed to imperative) language. XSL stylesheets have the advantage of being explicit, readable by humans and easy to debug and maintain. The program is generated in several steps, the final step being the transformation of the XML program description into the actual Scilab program. This final step may easily be adapted for other script languages such as Matlab (http:/www.mathworks.com) or Julia (http://julialang.org).

The generated program makes full use of Scilab's potential for improving the speed of calculations: vectorization and sparse matrices are used, and UMFPACK multi-frontal LU factorization \cite{davis:umfpack} is used for calculating the state and adjoint state. The optimization phase is performed using the Feasible Sequential Quadratic Programming (FSQP) algorithm  \cite{fsqp}, which is available as a Scilab module. Linearized statistics or Monte-Carlo simulations can be used to determine confidence intervals on estimated fluxes. Where Monte-Carlo simulations are used, Scilab allows for parallel execution of the optimization algorithm in a multi-core architecture.
\section{Numerical results}
\label{sec:num}
As an example, we considered the central metabolism of \emph{E. Coli}, described in \cite{millard2014sampling} and provided in the network examples in the \verb1influx_s1 distribution (the file \texttt{e\_coli.ftbl}). The experimental stationary data consist of mass spectrometry measurements of the intermediate metabolites Suc, ICit, PEP, PGA,  FruBP, Glc6P, Fru6P, Rib5P, Gnt6P, and of the extracellular flux of Acetate. For the adjoint vs. direct comparison in the stationary case, the original data were used. For the nonstationary case, synthetic noisy data were generated by a forward nonstationary simulation using the estimated values of flux from the stationary data and some realistic pool sizes (the file \texttt{e\_coli\_ns.fml} in the \texttt{sysmetab} distribution).

The results presented in this section were obtained on a dedicated Linux server hosting two ten-core Xeon E5-2660 v2 (2.20 GHz) processors with Scilab 5.5.2. Unless explicitly stated otherwise, computations used only one core of the processor, and every reported running time is the median over ten runs. Although the running times themselves may be very different if different hardware is used, we assume that the ratios between the different running times will remain similar.
\subsection{Adjoint versus direct gradient computation}
\subsubsection{Stationary data}
\label{sec:num:stat}
\begin{table}[tb]
\caption{\texttt{sysmetab} direct vs. adjoint average computation times and ratios for the reduced and full \emph{E. Coli} network with stationary data\label{tab:ratios}}
\begin{tabular}{llll}
 & Direct & Adjoint & Ratio\\
 \hline\\
Reduced network\\
 \hline
State computation \eqref{eq:cascade} & {1.5} & {1.5}& {1}\\
Cascade \eqref{eq:jac} vs. \eqref{eq:adjcascade} & {6}&  {0.35} & {17.5}\\
Gradient assembly \eqref{eq:fwdgrad} vs. \eqref{eq:adjgrad}   & {0.9}& {0.05} & {17.2}\\
Total & {8.4} & {1.9} & {4.47}\\
\\Full network\\
 \hline
State computation  \eqref{eq:cascade} & {3.5} & {3.5} & {1} \\
Cascade  \eqref{eq:jac} vs. \eqref{eq:adjcascade} & {47.2} &  {1.5}  & {31} \\
Gradient assembly \eqref{eq:fwdgrad} vs. \eqref{eq:adjgrad}   & {4.2} & {0.2}  & {19} \\
Total & {55} & {5.2}  & {10.6} \\
\hline\\
\scriptsize{Time unit=$10^{-3}$  s.}
\end{tabular}
\end{table}
Table \ref{tab:ratios} compares the average computation times from steps \eqref{eq:jac}-\eqref{eq:fwdgrad} in the direct approach with steps \eqref{eq:adjcascade}-\eqref{eq:adjgrad} in the adjoint approach.
We report the computation times required for solving the state equation so that the overall times of the two approaches may be compared. In Table \ref{tab:ratios}, time results from the reduced network (using the backwards tracing method \cite{antoniewicz2007}) of 1069 cumomers are contrasted with those obtained from the full original network of 5455 cumomers.

Since there are almost five times as many cumomers in the full network as in the reduced network, the time needed to compute the state derivative rises. In the direct approach, the contribution of this step to the overall time increases the ratio between the two approaches to approximately 10. In the adjoint approach, the size of the state derivative computation is in relation relation to overall time is smaller. The distribution of overall time reported in Table \ref{tab:ratios} clearly shows the efficiency of the adjoint computation in comparison with the direct computation.
\subsubsection{Nonstationary data}
\label{num:nonstat}
\begin{table}[tb]
\caption{\texttt{sysmetab} direct vs. adjoint average computation times and ratios for the reduced and full \emph{E. Coli} network with nonstationary data \label{tab:nonstat}}
\begin{tabular}{lrll}
 & Direct & Adjoint & Ratio\\
 \hline\\
Reduced network\\
 \hline
State computation \eqref{eq:discrcasc}& {21} & {21}& 1\\
Cascade \eqref{eq:discrdirect} vs. \eqref{eq:discradjoint}&{948} &  {15}& {60}\\
Gradient assembly  \eqref{eq:discrdirectgrad} vs. \eqref{eq:discradjgrad}& {11} & {16}& {0.70}\\
Total & {980} & {53} & {18.4}\\
\\Full network\\
 \hline
State computation  \eqref{eq:discrcasc} & {87} & {87} & {1} \\
Cascade \eqref{eq:discrdirect} vs. \eqref{eq:discradjoint}& {11214} &  {178}  & {63} \\
Gradient assembly  \eqref{eq:discrdirectgrad} vs. \eqref{eq:discradjgrad} & {118} & {90}  & {1.3} \\
Total & {11420} & {355}  & {32} \\
\hline\\
\scriptsize{Time unit=$10^{-3}$  s.}
\end{tabular}
\end{table}
Using the same network, we obtained estimated flux values from stationary data (given in the next section) and combined them with arbitrary pool size values comprising pool size measurements available in \cite{millard2014sampling}. The resulting values were used to simulate the time course of labeling states. We obtained the data by running the implicit ODE scheme \eqref{eq:discrcasc} up to a fixed terminal time $T$, and selecting the simulated measurements at time values ~$\tau_j, j=1,\ldots ,M$. Gaussian noise was then added by considering the standard deviation values of each of the original stationary labeling state data values.

The different time values comprise ten equally distributed values up to $\tau_M=10$ seconds, namely $\tau_j=j$ for $j=1,\ldots,10$.
We make use of the unconditionally stable property of the implicit Euler scheme in order to set the time step $h$ to $\tau_M/200$. Other experiments show that the outcome $x$  does not vary significantly for smaller values of $h$.

For the gradient computation, and more generally for the parameter estimations, we chose a value of $h$ larger than that used to generate the simulated data, typically $h=\tau_M/100$, so that the time discretization of the grid is not the same as for the forward problem.

Applying the same computation time analysis to these time-dependent data that we applied in the case of stationary MFA, we report the time distribution of the different steps of the gradient computation: state computation, cascade computation and gradient assembly, namely \eqref{eq:discrcasc}, \eqref{eq:discrdirect}, \eqref{eq:discrdirectgrad} for the time-dependent direct approach, and \eqref{eq:discrcasc}, \eqref{eq:discradjoint}, \eqref{eq:discradjgrad} for the time-dependent adjoint approach.

For the reduced \emph{E. Coli} network, associated computation times are presented in Table~\ref{tab:nonstat}. As expected, the time required to compute the state derivative using the direct method (row 2) is greater than the stationary cascade cost (see the second row of Table~\ref{tab:ratios}) multiplied by the overall number of time steps (i.e. 100). The cost of the state derivative (direct) cascade in the nonstationary case is in fact more than 150 times the cost of one stationary (direct) cascade. The ratio between the direct approach and the adjoint approach is 60, which demonstrates the efficiency of the adjoint approach. The ratio of over 18 for overall gain is seen to be significantly better than the ratios obtained in the stationary case: the adjoint approach benefits from the nonstationary case, while the direct gradient computation, as expected, is seen to be severely time consuming.
\subsection{Assessment of parameter estimates}
\label{sec:valid}
\subsubsection{Stationary data}
\begin{table}[tb]
\renewcommand{\arraystretch}{1.2}
\caption{Estimated free flux values and final RSS obtained by \texttt{influx\_s}, \texttt{sysmetab} and \texttt{13CFlux2}. \label{tab:compstat}}
\centering\begin{tabular}{llll}
\textbf{Software}	& \texttt{influx\_s}	& \texttt{sysmetab}	& \texttt{13CFlux2}\\
\hline\\
{Free fluxes} \\\hline
Glucupt\_1.n&  0.80651342	& 0.80651338 & 0.80634528\\
gnd.n	      &  0.14223488 & 0.14223479 & 0.13804339\\
out\_Ac.n   &  0.21300000 & 0.21300001 & 0.21299847\\
pyk.n       &  1.54425504 & 1.54425507 & 1.52116680\\
zwf.n       &  0.14233488 & 0.14233479 & 0.14145512\\
ald.x       &  0.72497526 & 0.72497650 & 0.89824047\\
eno.x       &  999        &  999        &  98486.1688\\
fum\_a.x    &  0.43135328 & 0.43136935 & 0.40973977\\
ppc.x       &  0.17317878 & 0.17317925 & 0.17122549\\
ta.x        &  0.56653486 & 0.56653403 & 0.56756226\\
tk1.x       &  0.17878854 & 0.17878863 & 0.17125450\\
tk2.x       &  0          & 0          & 0.00213\\
\\{Final RSS}\\\hline
&61.5141593 &	61.5141593 & 62.0877059\\
\end{tabular}
\end{table}
After comparing the adjoint and the direct approach in \verb1sysmetab1 we consider its validation by performing the complete estimation from stationary data. The provided \verb1e_coli.ftbl1 file was processed in \verb1influx_s1 with the \verb1--emu1 command line option. In addition, after conversion to the FML format, it was processed first with the last version of \verb$13CFlux2$ with default options and then in \verb1sysmetab1 with the command line options \verb$--reg=0$ (the default regularization term is not added to the RSS). Table~\ref{tab:compstat} presents the flux values obtained after optimization, (exchange fluxes given by \texttt{influx\_s} have been converted from normalized $[0,1[$ values to $[0,+\infty[$ interval)) and the final value of the RSS function $J(v)$.

As the exchange flux eno.x is nearly non-identifiable, the default upper bound is reached by \verb1sysmetab1 and \verb1influx_s1. We tried to add the constraint explicitly in the FML file before running \verb$13CFlux2$, but the optimization stopped with a very large value of the RSS. We therefore do not include \verb$13CFlux2$ in the discussion below, since we were not able to solve this problem.

As \verb1influx_s1 and \verb1sysmetab1 give very similar results, we compare their respective behavior in terms of overall computation time, which includes code generation, optimization  to obtain the optimal flux values, and linear statistics to obtain their confidence intervals. On the well-defined and quickly converging example of \verb1e_coli.ftbl1, the fastest is \verb1influx_s1 (6.7 vs. 11.9 seconds for \verb1sysmetab1). But
if we only consider optimization time, \verb1sysmetab1 is seen to be more efficient (0.35 vs. 0.74 seconds for \verb1influx_s1). Comparing sub-second results for a particular network may not be so informative, and we therefore considered a Monte-Carlo estimation of flux statistics with 1000 data samples, using all available cores of the processors. In this typical situation where the code has to be generated only once, \verb1influx_s1 took 105 seconds and \verb1sysmetab1 24 seconds.

We did the same Monte Carlo study involving 1000 samples on a poorly defined network describing the central metabolism of \emph{E. Coli} and reactions for amino acid biosynthesis (the file \verb$Ecoli.1.ftbl$, available as part of the \cite{influxs} additional material). Here, \texttt{influx\_s} shows its superiority in terms of numerical stability. But even though \verb1sysmetab1 requires more (but less costly) iterations than \texttt{influx\_s} for each optimization, it is faster than \texttt{influx\_s}, terminating in 110 seconds instead of 480.
\subsubsection{Nonstationary data}
\begin{figure}[t]
\sf\footnotesize \noindent\begin{tabularx}{\linewidth}{cc}
~~~Suc\\\includegraphics[width=.45\linewidth]{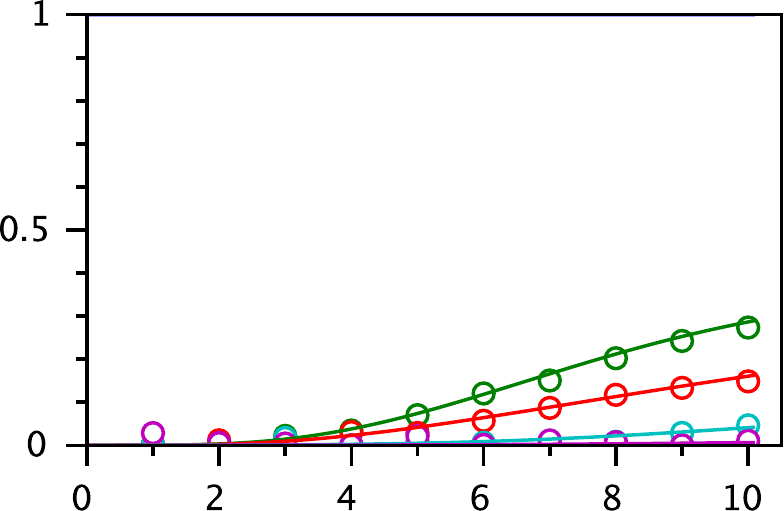} & \includegraphics[width=.45\linewidth]{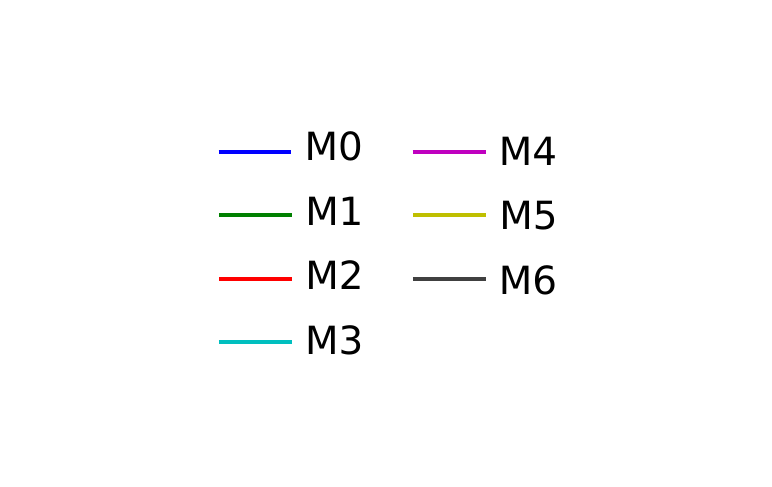} \\
~~~ICit & ~~~PEP \\
\includegraphics[width=.45\linewidth]{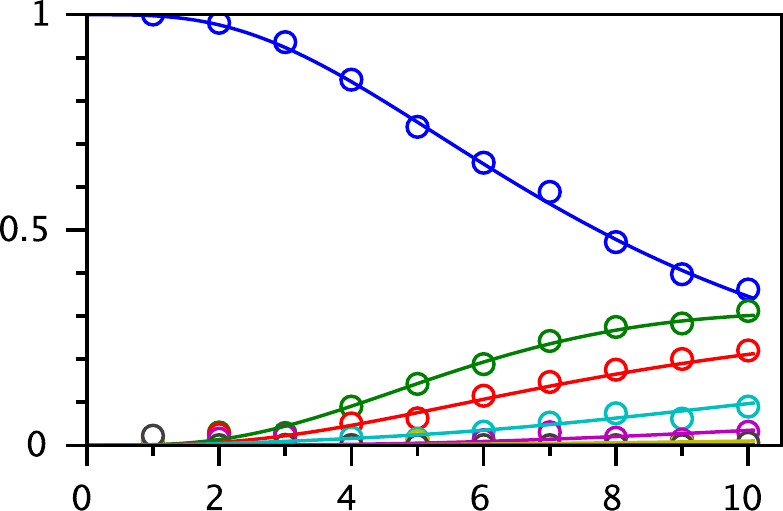} & \includegraphics[width=.45\linewidth]{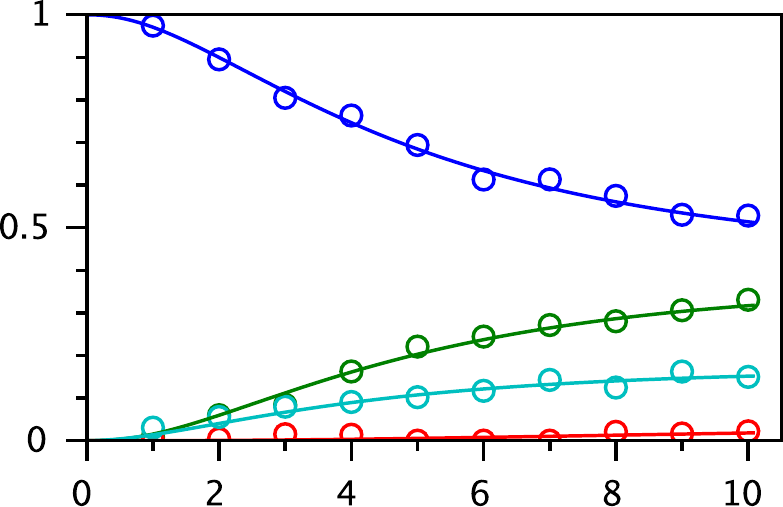}\\
~~~PGA & ~~~FruBP\\
\includegraphics[width=.45\linewidth]{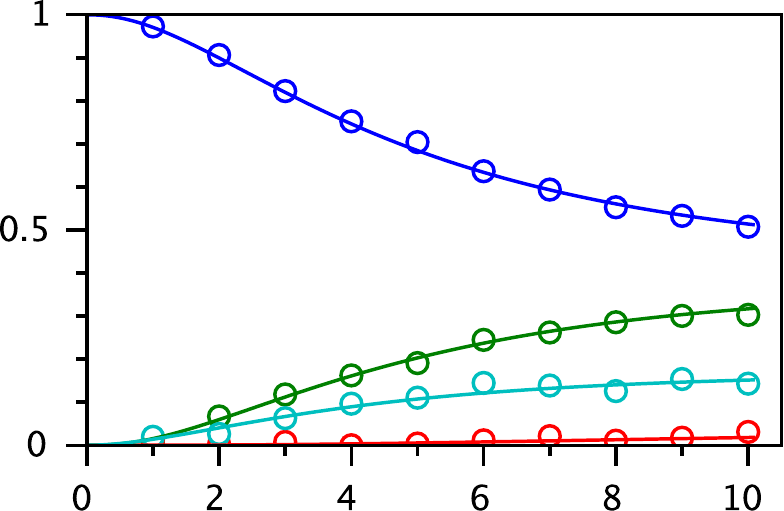} & \includegraphics[width=.45\linewidth]{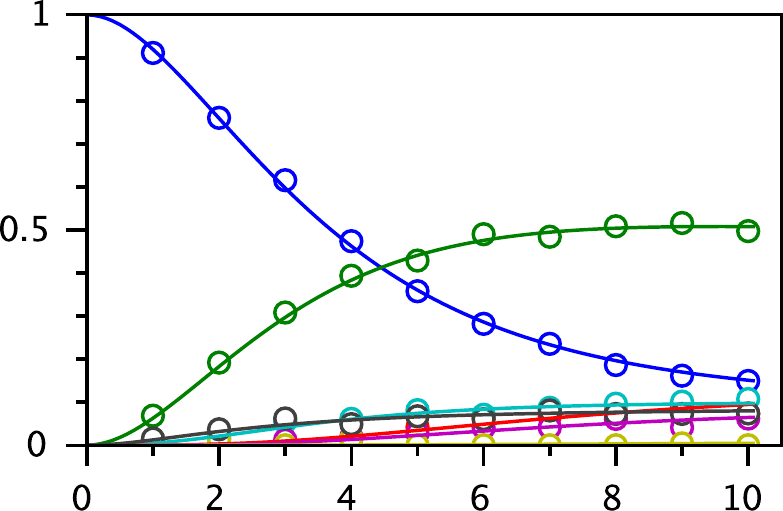}\\
~~~Glc6P & ~~~Fru6P\\
\includegraphics[width=.45\linewidth]{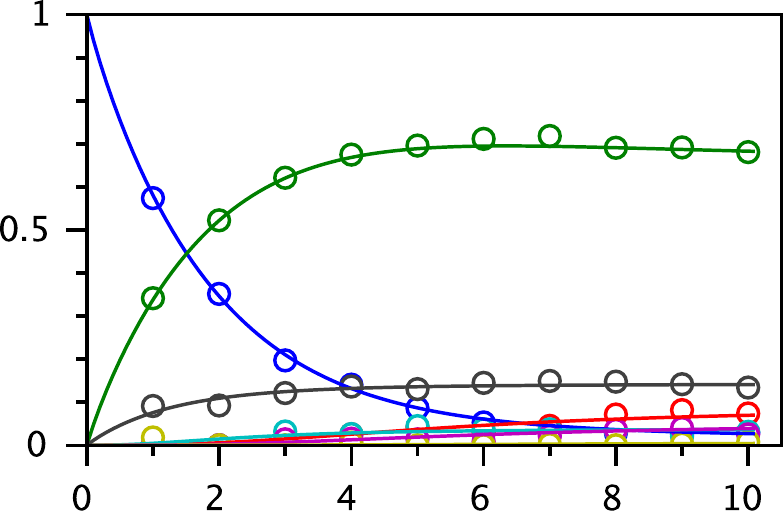} & \includegraphics[width=.45\linewidth]{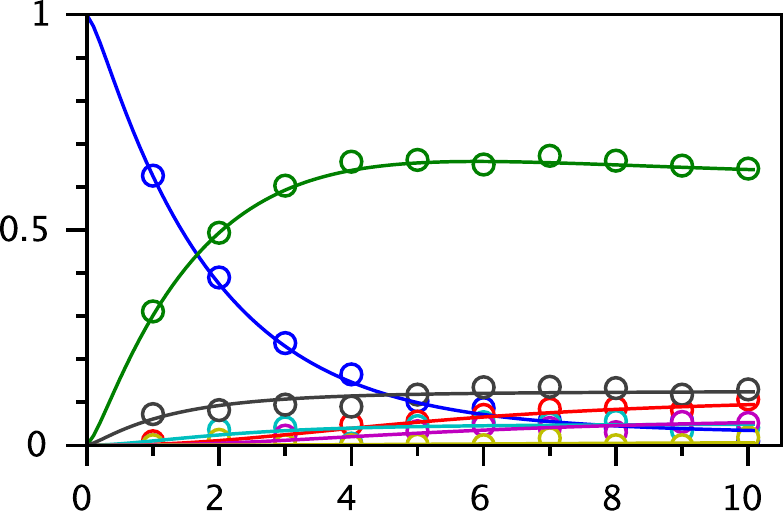}\\
~~~Rib5P & ~~~Gnt6P\\
\includegraphics[width=.45\linewidth]{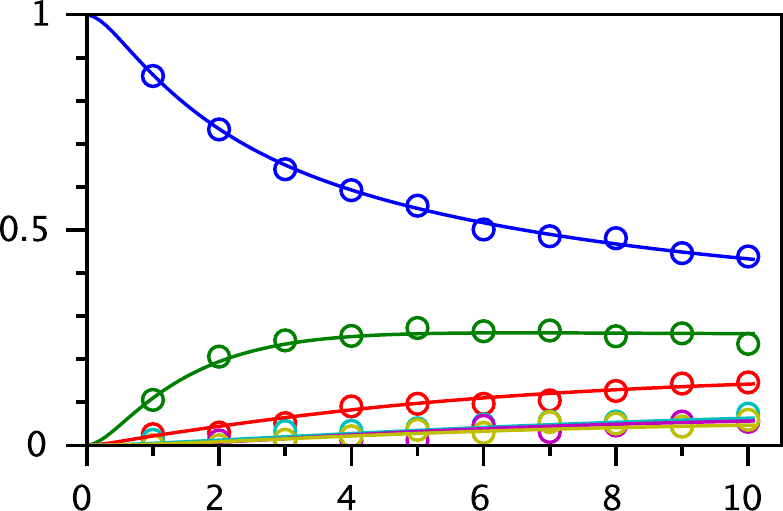} & \includegraphics[width=.45\linewidth]{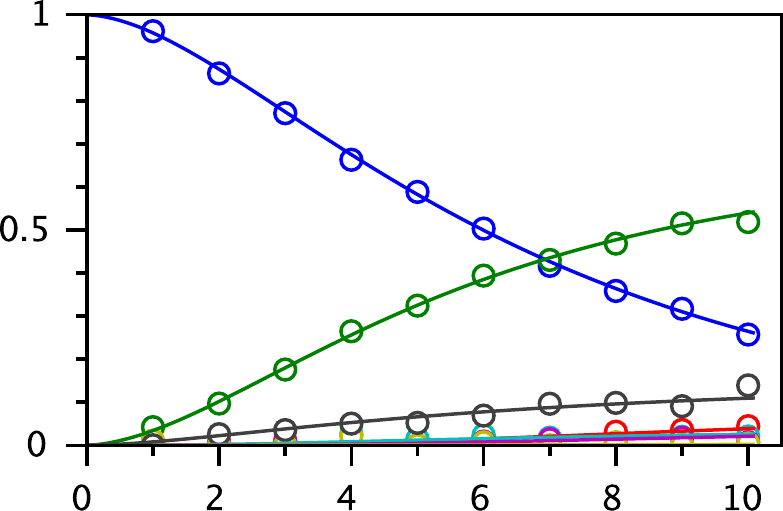}\\
~~~~Time (s) & ~~~~Time (s)\\\end{tabularx}
\caption{Experimental mass isotopomers fractions (dots) reconstruction by simulated data (solid lines) after optimization of the parameters.}
\label{fig:meas}
\end{figure}
We are first interested in checking wether the time-dependent labeling measurements are best fitted. Figure~\ref{fig:meas} shows the reconstruction of mass isotopomer fractions of measured metabolites using the free fluxes and pool size values obtained after optimization. Comparing these simulated data with the different time measurements marked with circles demonstrates that synthetic data are well recovered.
The estimated pool sizes and the estimated flux values obtained from the nonstationary data are reported in Table~\ref{tab:flux} and contrasted with the fluxes obtained from the stationary data. The 95\% confidence intervals were computed from the empirical repartition of optimal parameters, estimated by the Monte Carlo method (1000 resamplings of stationary and nonstationary data were used).
\begin{table}[tb]
\renewcommand{\arraystretch}{1}
\caption{Comparison of estimated parameter values by \texttt{sysmetab} from stationary and nonstationary data \label{tab:flux}}
\centering\begin{tabular}{lllll}
\textbf{Data}	& \multicolumn{2}{c}{Nonstationary}	& \multicolumn{2}{c}{Stationary} \\
\hline\\
{Free fluxes}& Median & 95\% C.I.  & Median  & 95\% C.I. \\ \hline
Glucupt\_1.n & 0.807 & [0.803, 0.811] & 0.806 & [0.798, 0.813] \\
pyk.n        & 1.522 & [1.293, 1.547] & 1.535 & [1.355, 1.552] \\
zwf.n        & 0.150 & [0.131, 0.182] & 0.143 & [0.114, 0.177] \\
gnd.n	     & 0.143 & [0.128, 0.160] & 0.136 & [0.101, 0.169] \\
out\_Ac.n    & 0.213 & [0.213, 0.213] & 0.213 & [0.213, 0.213] \\
ald.x        & 0.647 & [0.579, 0.717] & 0.731 & [0.558, 0.929] \\
eno.x        & 999  &  [2.371, 999]   & 88.61 & [0.638, 194.2] \\
ta.x         & 0.570 & [0.531, 0.611] & 0.556 & [0.486, 0.635] \\
tk1.x        & 0.183 & [0.151, 0.214] & 0.157 & [0.086, 0.222] \\
tk2.x        & 0.000 & [0.000, 0.021]  & 0.005 & [0.000, 0.066] \\
fum\_a.x     & 0.100 & [0.000, 0.480] & 0.310 & [0.000, 66.50] \\
ppc.x        & 0.241 & [0.080, 0.427]& 0.189 & [0.083, 0.363] \\
\\{Free pool sizes} & Median & 95\% C.I. \\\hline
\bf FruBP & 1.860 & [1.857, 1.862]\\
\bf Glc6P & 1.435 & [1.428, 1.441] \\
\bf Fru6P & 0.425&  [0.420, 0.431]\\
\bf Rib5P& 0.020 &  [0.013, 0.026]\\
\bf GA3P & 0.470 &  [0.470, 0.470]\\
\bf PEP & 0.110 &   [0.110, 0.110]\\
\bf Sed7P & 0.057 & [0.052, 0.063]\\
      Suc & 0.854 & [0.078, 2.427]\\
      ICit& 1.281 & [0.376, 1.707]\\
      PGA & 0.478 & [0.086, 0.868]\\
    Gnt6P & 0.836 & [0.736, 1.022]\\
      AKG & 0.638 & [0.000, 1.786]\\
    Ery4P & 0.278 & [0.004, 0.619]\\
      OAA & 0.000 & [0.000, 1.501]\\
      Pyr & 0.530 & [0.000, 2.756]\\
\end{tabular}
\end{table}
We first note that the estimated values obtained from time-dependent labeling state data are similar to those obtained from the stationary state. Then, focusing on the confidence intervals of fluxes, we note that for almost all of them, considering ten measurements during the transient phase gives better accuracy than using a single measurement when the stationary state is reached. As far as the pool sizes are concerned, for those which have an associated measurement (in bold font), the measurement is well recovered, but apart from Gnt6P all the pools have a significantly larger confidence interval. The choice of measurement times could be further optimized in order to improve the parameter statistics.

If we now compare the computation time on our hardware for the two approaches, we remark that estimating the fluxes and pool size parameters takes almost 8 minutes using the direct method. Using the adjoint method this estimation takes only 36 seconds (21 seconds just for the optimization) and the computation of confidence intervals (Monte Carlo method with 1000 samples) takes 13 minutes, which demonstrates the efficiency of the approach.
\section{Conclusion}
In this paper we address the MFA problem by making use of the adjoint state, in the spirit of a general optimization problem governed by a state equation, as in control theory. The proposed methods and algorithms are included in the \verb1sysmetab1 software, distributed under an Open Source license. In the stationary case, results demonstrate the efficiency of the approach in terms of precision and computation time when a typical metabolic network is considered. Estimated values were validated against those obtained by other reference software. With the adjoint framework, only vectors (instead of matrices) are updated, which considerably speeds up the computation of the gradient. In the nonstationary case, we considered the same network with synthetic noisy data. On our hardware and using 100 time steps of the ODE integration scheme, a forward simulation followed by the adjoint gradient computation takes 53 milliseconds, instead of almost 1 second when the gradient is computed with the direct method, and a full estimation of fluxes and pool sizes takes 36 seconds. These timings make the nonstationary MFA problem computationally tractable for larger networks and make \verb1sysmetab1 competitive with  other software.

Although \verb1sysmetab1 does not yet implement it, the EMU framework can be used simultaneously with the adjoint approach, and this simultaneous implementation could represent the best computational solution when only MS measurements are used. Although this approach favors gradient-based methods, it can also be used in Gauss-Newton type methods: in this case it allows the computation of the output sensitivity matrix at the cost of $n_y$ adjoint gradients, where $n_y$ is the size of the measurement vector at a given time \cite{sandu}.

Our current development efforts include improving the speed of the code generation step, and implementing high-order ODE solvers and alternative methods for computing nonlinear confidence intervals \cite{beale}.
\section*{Acknowledgments}
This work was performed, in partnership with the SAS PIVERT, within the frame of the French Institute for the Energy Transition (Institut pour la Transition Energ\'etique (ITE) P.I.V.E.R.T. (www.institut-pivert.com) selected as an Investment for the Future ("Investissements d'Avenir"). This work was supported, as part of the Investments for the Future, by the French Government under the reference ANR-001
%

%
\begin{IEEEbiography}{St\'ephane Mottelet}
received a Master's degree in Computer Science and
Engineering, as well as a Master's degree and a Ph.D. in Control Engineering (1994),
from University of Technology of Compi\`egne (UTC, Compi\`egne,
France). Since 1995 he has been an assistant professor in the Applied
Mathematics Laboratory of UTC and joined in 2015 the Process Engineering Laboratory in the same university. His current research interests include inverse
problems, optimization, control theory, partial differential equations and their applications to systems biology and bioprocess engineering.
\end{IEEEbiography}
\begin{IEEEbiography}{Gil Gaullier}
received a Master's degree in Computer Science, as well as a MSc in Numerical Analysis from Pierre \& Marie Curie University (Paris, France) and a Ph.D. in Image Processing (2013) from Strasbourg University (France). He is currently a Post-Doctoral Associate in the Applied Mathematics Laboratory of UTC. His fields of interest include inverse problem and optimization for interdisciplinary applications such as computational biology.
\end{IEEEbiography}
\begin{IEEEbiography}{Georges Sadaka}

\end{IEEEbiography}
\end{document}